\documentclass[fleqn,usenatbib]{mnras}

\usepackage[T1]{fontenc}
\usepackage{ae,aecompl}

\usepackage{graphicx}	
\usepackage{amsmath}	
\usepackage{amssymb}	
\usepackage{courier}

\title[Phase Space Spirals in GDR2]{Emergence of the {\it Gaia} Phase Space
  Spirals from Bending Waves}

\author[Keir Darling et al.]{
Keir Darling\thanks{E-mail: 13kd39@queensu.ca}
and Lawrence M. Widrow\thanks{E-mail: widrow@queensu.ca}
\\
Department of Physics, Engineering Physics \& Astronomy, Queen's University, Stirling Hall, Kingston, ON K7L 3N6, Canada
}

\date{Accepted XXX. Received YYY; in original form ZZZ}

\pubyear{2018}

\begin{document}

\label{firstpage}
\pagerange{\pageref{firstpage}--\pageref{lastpage}}
\maketitle

\begin{abstract}
	We discuss the physical mechanism by which pure vertical bending waves in a stellar disc evolve to form phase space spirals similar to those discovered by Antoja et al. in {\it Gaia} Data Release 2.  These spirals are found by projecting Solar Neighbourhood stars onto the $z-v_z$ plane. Faint spirals appear in the number density of stars projected onto the $z-v_z$ plane, which can be explained by a simple model for phase wrapping.  More prominent spirals are seen when bins across the $z-v_z$ plane are coloured by median $v_R$ or $v_\phi$. We use both toy model and fully self-consistent simulations to show that the spirals develop naturally from vertical bending oscillations of a stellar disc.  The underlying physics follows from the observation that the vertical energy of a star (essentially, its ``radius'' in the $z-v_z$ plane) correlates with its angular momentum or, alternatively, guiding radius.  Moreover, at fixed physical radius, the guiding radius determines the azimuthal velocity.  Together, these properties imply a link between in-plane and vertical motion that lead directly to the {\it Gaia} spirals.  We show that the cubic $R-z$ coupling term in the effective potential is crucial for understanding the morphology of the spirals.  This suggests that phase space spirals might be a powerful probe of the Galactic potential.  In addition, we argue that self-gravity is necessary to properly model the evolution of the bending waves and their attendant phase space spirals.
\end{abstract}



\section{INTRODUCTION}

{\it Gaia} Data Release 2 (GDR2) \citep{brown2018}, which includes
measurements of the six-dimensional phase space coordinates for some
seven million stars, has afforded astronomers an unprecedented picture
of the local stellar kinematics in our Galaxy \citet{katz2018}.
Arguably, the most intriguing result from the nascent analysis of this
data set has been the discovery of spiral patterns in certain phase
space projections of Solar Neighborhood stars by \citet{antoja2018}.
They selected stars in a circular arc that spanned $8^{\circ}$ in
Galactic azimuth $\phi$ and a range in Galactocentric radius $R$ from
$8.24\ {\rm kpc}$ to $8.44\ {\rm kpc}$.  They then computed the number
density distribution of the $\sim$900k stars within this region as a
function $z$ and $v_z$, the position and velocity in the direction
perpendicular to the Galactic midplane.  The resulting plot showed a
spiral pattern of 1-2 complete wraps within a region of the $z-v_z$
plane extending to $\sim 700\,{\rm pc}$ in $|z|$ and
$\sim 40\,{\rm km\,s^{-1}}$ in $|v_z|$.  More prominent spirals
appeared when they ``coloured'' the $z-v_z$ plane by median $v_R$ and
$v_\phi$.  \citet{antoja2018} interpreted the spirals as a telltale
sign of phase wrapping and were able to explain their general
morphology by a simple model for the local gravitational potential.

In this paper, we delve deeper into the physics of the ``{\it Gaia}
spirals''.  Our hypothesis is that a pure bending wave naturally
evolves to form spiral patterns in the $\left
(z,\,v_z,\,v_\phi,\,v_R\right )$ phase space.  (For a different take
on some of the same ideas, see \citet{binney2018}.)  Bending waves are
a natural feature of disc galaxies.  The most conspicuous examples are
the warps seen in edge-on galaxies (See \citet{binney1992} and
\citet{sellwood2013} for reviews of galactic warps.).  These warps
amount to a bending of both HI and stellar discs from the midplane by
$\sim 1-3\,{\rm kpc}$.  Recently, \cite{xu2015} found asymmetries in
the number counts of stars above and below the disc at Galactocentric
radii between $12\,{\rm kpc}$ and $18\,{\rm kpc}$, which they
interpreted as evidence for ripples or corrugations in the disc. In
addition, \citet{schonrich2015} found wavelike patterns in the mean
vertical motion of Solar Neighbourhood stars as a function of $L_z$,
the angular momentum about the rotation axis of the Galaxy.  Since
$L_z$ is roughly proportional to guiding radius (at least, where the
rotation curve is approximately flat) these velocity ripples may well
be the velocity counterpart to those seen in number counts
\citep{chequers2018}.  Further evidence for bending and breathing
modes has been seen in both number counts and bulk motions of Solar
Neighbourhood stars (see, for example, \citet{widrow2012,
  williams2013, carlin2013, yanny2013, brown2018, bennett2018}).  They
can be excited by a passing satellite \citep{widrow2012, feldman2015,
  gomez2013, gomez2017}, spiral structure or the bar
\citep{debattista2014, monari2015}, or even shot noise in an N-body
simulation \citep{chequers2017}.

In this paper we sidestep the issue of what perturbs the disc by
imposing an initial {\it ad hoc} vertical perturbation.  The excitation of
bending and breathing waves due to a passing satellite is discussed
in detail by \citet{widrow2014} and demonstrated in N-body simulations
of both isolated discs in discs in fully cosmological simulations by
\citet{gomez2013, feldman2015, gomez2017, chequers2018}.  In essence, these
exitations are the early stages of the disc-heating events discussed
in \citet{toth1992} and \citet{sellwood1998}.  The excitation of
perturbations in the context of phase space spirals using the impulse
approximation is discussed in \citet{binney2018}.

In Section 2 we consider a series of toy-model simulations where test
particles are evolved in time-dependent, axisymmetric potentials.
These simulations allow us to illustrate the essential physics of
phase space spirals and, in particular, highlight the importance of the
non-separable nature of the effective potential for disc stars.  In
Section 3, we then demonstrate the link between bending waves and
phase space spirals in a fully self-consistent disc-bulge-halo model
of the Milky Way using standard N-body methods.  We conclude in
Section 4 with a summary of our results and suggestions for further
investigations of these ideas.

\section{KINEMATIC SPIRALS}\label{KinematicSpirals}

\subsection{Distribution Function}\label{DF}

Consider a distribution of test particles in a time-independent,
axisymmetric potential $\Phi\left (R,\,z\right )$.  All particle orbits
admit two integrals of motion: the angular momentum about the 
symmetry axis $L_z = Rv_\phi$ and the total energy $E=\frac{1}{2}v^2 +
\Phi\left (R,\,z\right )$.  If the potential is an additively
separable function, that is, if it can be written in form

\begin{equation}\label{seppot}
\Phi\left (R,\,z\right ) = \psi(R) + \chi(z)~,
\end{equation}

\noindent then the vertical energy

\begin{equation}
E_z = \frac{1}{2}v_z^2+\Phi(R,z) - \Phi(R,0)
\end{equation}

\noindent is also an integral of motion.  In this case, it is
convenient to define the planar energy $E_p = E - E_z$.  Even if the
potential cannot be written in the form given by Eq.\,\ref{seppot},
the vertical energy is approximately conserved for particles on nearly
circular orbits \citep{binney2008}.

Analytic functions of $E_p$, $E_z$, and $L_z$ can be used as the
building blocks for a disc that is close to equilibrium.  Following
\citet{kuijken1995} we consider the quasi-isothermal distribution
function (DF)


\begin{equation}\label{DFeq}
f(E_z,E_p,L_z) = \frac{g}{\sigma_z\sigma_R^2}  
\exp\bigg[ - \frac{E_p - E_c}{\sigma_R^2}-\frac{E_z}{\sigma_z^2}\bigg]
\end{equation}

\noindent where $E_c$ is the energy of a particle on a circular orbit
with angular momentum $L_z$, and $ g $, $ \sigma_R $, $ \sigma_z $, and $ E_c $ are functions of $ L_z $.

The function $g(L_z ) $ determines the radial profile of the
surface density, which is obtained by integrating the DF over
velocities and $z$.  Furthermore, the guiding radius $R_c$ for a
particle on a circular orbit is an implicit function of $L_z$.  Thus,
we can use $R_c$, which is an integral of motion, as a proxy for
radius $R$ when constructing the DF.  For example, an exponential disc
is obtained by choosing $g\propto \exp{(-R_c/R_d)}$.

Our interest here is in stars in the Solar Neighbourhood.  We are
therefore led to consider the idealized case where

\begin{equation}\label{simpleg}
g(L_z)\propto\delta(L_z-L_{z0}).
\end{equation} 

\noindent The density is then given by

\begin{equation}
\rho(R,\,z) \propto \exp\bigg[-\frac{ \Phi_{\rm eff}\left (R,\,0\right ) -
    E_c}{\sigma_R^2} -\frac{\Phi_z\left (R,\,z\right )}{\sigma_z^2}\bigg]
\end{equation}

\noindent where $\Phi_{\rm eff}(R) = \Phi(R,\,0) + L_{z0}^2/2R^2$ is
the effective potential for a particle with angular momentum $L_{z0}$
and
$\Phi_z\left (R,\,z\right ) \equiv \Phi_(R,\,z) - \Phi\left
  (R,\,0\right )$.
Note that the peak of the radial distribution of particles coincides
with the minimum of the effective potential, that is, the guiding
radius of a particle with angular momentum $L_{z0}$.  The main
advantage of this DF is that it easily sampled.  It is, however, too
simplistic for our purposes as we now describe.

In the DF given by Eq.\,\ref{DFeq}, the angular momentum determines a
particle's guiding radius and hence the probability distribution functions
for a particle's phase space coordinates $R,\,z,\,v_R,\,v_\phi,$ and
$v_z$.  If we introduce a bending perturbation to this DF, that is, if 
we displace the DF in either $z$ or $v_z$, then the distribution will
phase wrap in the $z-v_z$ plane producing a spiral pattern similar to the one seen
in the number counts.  However, since the DF is separable in $z-v_z$
and $R-v_R-v_\phi$, spiral patterns will not appear when the 
$z-v_z$ plane is coloured by median $v_R$ or $v_\phi$.  

Of course, in a real galaxy one has a distribution of stars in $L_z$
at a given $R$.  To mimic this effect while retaining the simplicity of
Eq.\,\ref{simpleg} we consider a superposition of two delta-function
distributions,

\begin{equation}\label{g}
g(L_z) = \alpha_1\delta(L_z-L_{z1}) + \alpha_2\delta(L_z - L_{z2})~,
\end{equation}

\noindent which we refer to as populations 1 and 2. In what follows,
we assume that $ \alpha_1 = \alpha_2 = 0.5 $ and choose $ L_{z1} $ and
$ L_{z2} $ so that the two guiding radii are $ 7.41 $ kpc and $ 7.94 $
kpc, respectively.  Furthermore, we assume that $\sigma_R$ and
$\sigma_z$ are given by

\begin{equation}
\begin{split}
\sigma_R(R_c) & = \sigma_{R,0}e^{-\left (R_c - R_s\right )/2R_d}\\
\sigma_z(R_c) &= \sigma_{z,0}e^{-\left (R_c - R_s\right )/2R_d}
\end{split}
\end{equation}
 with $ R_d = 2.1 $ kpc,
$ \sigma_{R,0} = 26\ \text{km}\,\text{s}^{-1}$, and
$ \sigma_{z,0} = 16\ \text{km}\,\text{s}^{-1}$.

\subsection{$ R-z $ Coupling in the Potential}\label{ToyModel}

In this section we consider a toy model potential that allows us to
isolate the effect $R-z$ coupling terms in the effective potential
have on phase space spirals.  Recall that in the epicycle
approximation \citep{binney2008} one expands the effective potential
to terms quadratic in $R-R_c$ and $z$.  To the extent that the
approximation holds, stars execute simple harmonic motion in $R$ and
$z$.  Of course, without the higher order terms in the potential, all
stars would orbit in the $z-v_z$ plane at the same frequency and phase
space spirals would never develop.

As discussed above, the vertical energy $E_z$ is conserved so long as
the potential is additively separable.  We first consider a model for
the potential that is additively separable but includes higher order
terms in $z$ and $R-R_c$.  As we'll see, spirals appear in this model
but lack many of the qualitative features seen in the data.  We then
augment this model with an $\left (R-R_c\right )z^2$ term.  This term,
along with an $\left (R-R_c\right )^3$ term, are the leading
corrections to the epicycle approximation.  The spirals in this case
have a richer phenomenology than the ones in our separable model and
and share many of the features with the spirals seen in {\it Gaia} DR2.

\subsubsection{kinematic spirals in an additively separable
  potential}\label{uncoupled}

Consider an additively separable potential given by Eq.\,\ref{seppot}
with

\begin{equation}
\chi(z) = ab^2\sqrt{\frac{z^2}{b^2}+1} + cz^2
\end{equation}

\noindent and 

\begin{equation}
\psi(R) = \frac{v_o^2}{2}\ln{\left (R^2 + R_o^2\right)}~.
\end{equation}

\noindent The form of the vertical potential $\chi$ is similar to that from the Oort
problem analysis of \citet{kuijken1989} while $\psi$ is the
logarithmic potential (see, for example, \citet{binney2008}), which
gives a flat rotation curve for $R\gg R_0$. In what follows we choose 
$ a = 0.07\ \text{Myr}^{-2} $, $ b = 23 $ pc and $ c= 0.006\ \text{Myr}^{-2} $,
which give a vertical potential consistent with models for the vertical potential
in the Solar Neighborhood (See \citet{widrow2014}.). We also choose 
$ v_o = 217\ \text{km}\text{s}^{-1} $ and $ R_o = 6.1 $ kpc to give a radial potential 
consistent with the realistic Milky Way model considered in Section~\ref{MWP}. 

\begin{figure}
	\centering  
	\includegraphics[width=7cm]{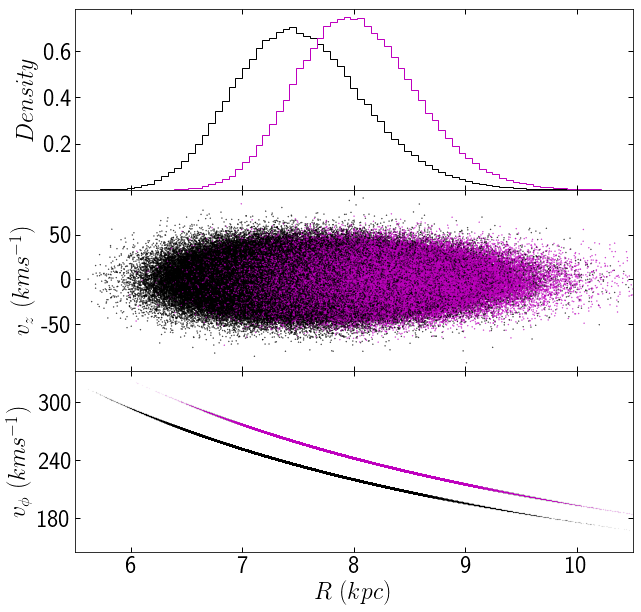}
	\caption{Properties of the initial conditions for the
          test-particle simulation described in Section\,\ref{uncoupled}.
In the top panel we show a normalized histogram of the particle number 
as a function of $R$ for populations 1 (black) and 2 (magenta), with a bin size of $ \Delta R = 0.06 $ kpc. In the middle panel
we show the distributions of stars in the two populations as a function
of $R$ and $v_z$ while in the bottom panel we show distributions as a
function of $R$ and $v_\phi$.}\label{ICplots}
\end{figure}

\begin{figure}
	\centering  
	\includegraphics[width=5cm]{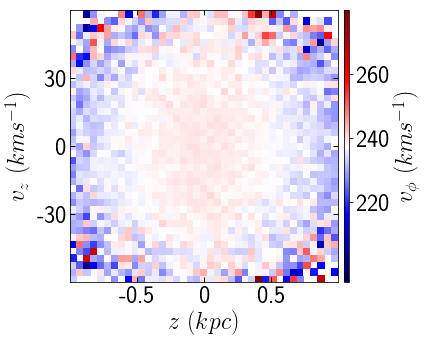}
	\caption{The $z-v_z$ plane colored by mean $v_\phi$ for the
          simulation described in Section\,\ref{uncoupled}.  The bins
          have size $ \Delta z = 0.05 $ kpc, and
          $ \Delta v_z = 3 \ \text{km}\text{s}^{-1} $.  Bins with
          fewer than 10 particles are ignored.}\label{ICvphi}
\end{figure}

We next sample the DF for 500k particles.  Properties of the initial
conditions are exhibited in Fig.~\ref{ICplots} and \ref{ICvphi}.  The
top panel of Fig.~\ref{ICplots} shows the number density of stars as a
function of Galactocentric radius for the two distributions.
Evidently, the model has the desired property that the distribution of
stars at a given radius comprises an admixture from the two $L_z$
populations. The bottom panel shows the distribution of
stars as a function of $R$ and $\mathbf{v_\phi}$.  Since $v_\phi = L_z/R$, the
distribution at a given radius is essentially the sum of two delta
functions with population 1 stars having a lower azimuthal velocity
than the population 2 stars at the same $R$. Finally, in the middle
panel, we show the distribution in $ R $ and $v_z$.  We note that stars at smaller
$R$ preferentially come from the higher $L_z$ distribution and have a
higher $\sigma_z$.  
  
Fig.~\ref{ICvphi} shows the $z-v_z$ plane coloured by mean $v_\phi$.  
Stars at large radii in the $z-v_z$ plane, that is, stars with high  
$E_z$ preferrentially come from stars in population 1, which have  
lower $v_\phi$, as seen in the bottom panel of Fig.~\ref{ICplots}.  

\begin{figure}
	\centering 
	\includegraphics[width=4.5cm]{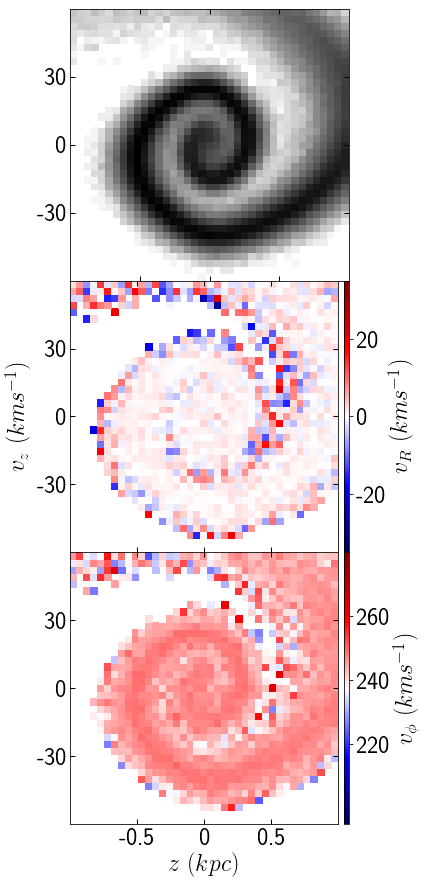}
	\caption{Number density and mean radial and azimuthal
          velocities in the $z-v_z$ plane for the toy model considered
          in Section \,\ref{uncoupled} (toy model with additively
          separable potential).  Top panel shows the number
          density.  Bins in the middle and lower panels are coloured
          by $v_R$ and $v_\phi$, respectively.  Bin sizes and particle
          number threshold are the same as in
          Fig.\,\ref{ICvphi}.}\label{ToyModelNoCoupling}
\end{figure}

The initial DF shown in Fig.\,\ref{ICvphi} is perturbed by displacing
it by $30\,{\rm km\,s^{-1}}$ in $v_z$.  The particles are then evolved
using a standard leap-frog algorithm for $300\,{\rm Myr}$ and the
results are shown in Fig.\,\ref{ToyModelNoCoupling}.  The number
density shows the spiral pattern typical of phase wrapping.  Stars
orbit the $z-v_z$ plane in a clockwise sense and since those at large
$E_z$ have a lower vertical epicycle frequency, the DF quickly
develops a trailing spiral.  Also shown is the $z-v_z$ plane coloured
by mean $v_R$ and $v_\phi$.  Here, the pattern is essentially
featureless, apart from the imprint of the number density
distribution: With our separable potential the in-plane and vertical epicyclic
motions completely decouple.

\subsubsection{kinematic spirals in a coupled potential}\label{coupled}

Next we consider the addition of a $R-z$ term to the toy model 
potential:

\begin{equation}\label{CoupledEq}
\Phi(R,z) = \psi(R) + \chi(z) + \xi(R-R_s)z^2 ~.
\end{equation}

\noindent $ R_s $ is a constant radius characteristic of the distribution 
(If a Taylor expansion of the effective potential is performed for 
each $ L_z $ population, $ R_s $ is replaced by $ R_c(L_z) $). For illustrative 
purposes, we choose
$\xi = -0.24\ {\rm Myr}^{-2} {\rm kpc}^{-1}$ and $ R_s = 8\ {\rm kpc} $.
These values are consistent with the cubic terms that arise in
realistic Milky Way potentials.  The main effect of the new term is to
introduce a linear dependence in $R$ to the vertical epicyclic
frequency. In particular the difference between the vertical
frequency for a population 1 star with $R_c = 7.41\,\rm kpc$ and a
population 2 star with $R_c = 7.94\,{\rm kpc}$ is $0.25\ {\rm Myr^{-1}}$.

\begin{figure}
	\centering 
	\includegraphics[width=4.5cm]{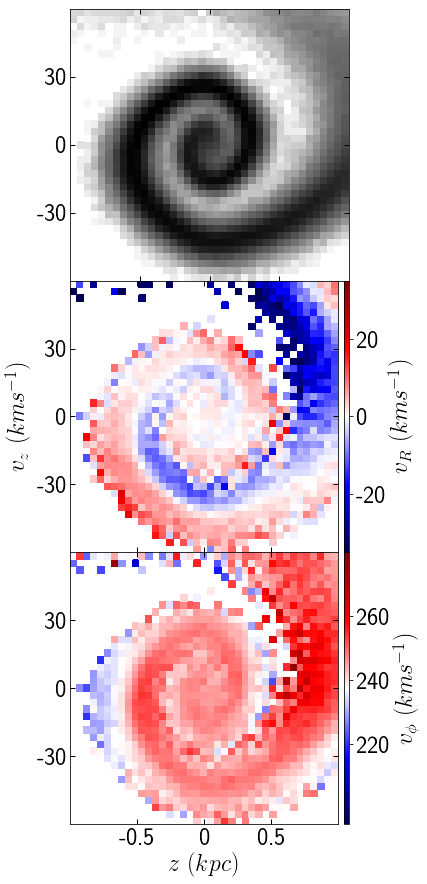}
	\caption{Same as Fig.\,\ref{ToyModelNoCoupling} but for the
          model considered in Section \,\ref{coupled} (toy model with
          $R-z$ coupling included in the
          potential).}\label{ToyModelCoupling}
\end{figure}

As before we sample an initial distribution (Eq.~\ref{DFeq}) with 500k
particles, perturb it and integrate the perturbed distribution for
$300\, {\rm Myr}$.  The results are shown in
Figure~\ref{ToyModelCoupling}.  Not surprisingly, the spirals that
appear in the $z-v_z$ number density distribution are virtually
identical to the ones that appear in our separable model.  However,
the $v_R$ and $v_\phi$ spirals are qualitatively different.  In
particular, there are gradients across the spiral arms and along their
inner and outer edges.  The addition of the $(R-R_s)z^2$ term implies
a coupling between the in-plane and vertical epicyclic motions.

\subsection{Realistic Milky Way Potential}\label{MWP}
 
\begin{figure}
	\centering 
	\includegraphics[width=7cm]{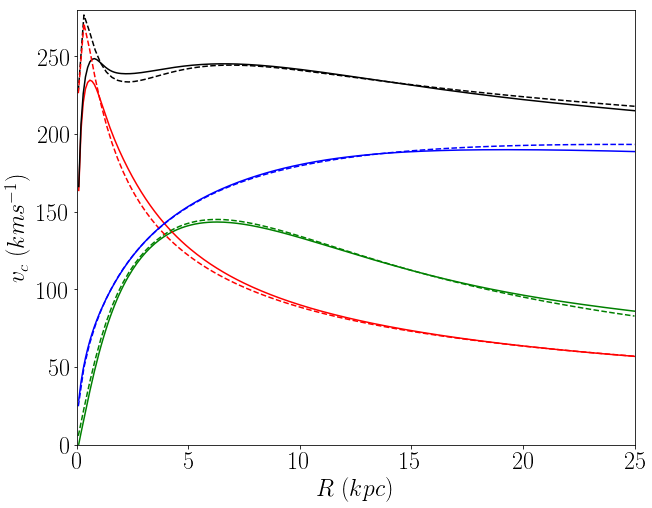}
	\caption{Circular speed curve for the Milky Way potential
          discussed in Section\,\ref{MWP} (dashed curves) and
          self-consistent model discussed in
          Section\,\ref{selfgravity} (solid curves).  For each model we
          show the total circular speed (black curves) and
          contributions from the bulge (red) disk (green) and halo
          (blue). }\label{vcirc}
\end{figure}

We conclude this section on test-particle simulations by considering a
realistic Milky Way potential that comprises an exponential disc, an NFW
halo \citep{navarro1997}, and a Hernquist bulge \citep{hernquist1990}.
We approximate the disc as a superposition of three Miyamoto-Nagia
potentials \citep{miyamoto1975, smith2015}.  The potential is similar
to the \texttt{MWPotential2014} model included with the \textsc{Galpy} code \citep{bovy2015}, but the specific model we used can be defined in \textsc{Galpy} using \texttt{HernquistPotential(a=0.035, amp=0.5)}, \texttt{MN3ExponentialDiskPotential(amp=7, hr=2.8/8, hz=0.3/8, sech=True)} and \texttt{NFWPotential(amp=5, a=1.4)}. The decomposition of the circular speed curve into the three
components is shown in Fig.~\ref{vcirc}.

\begin{figure}
	\centering 
	\includegraphics[width=4.5cm]{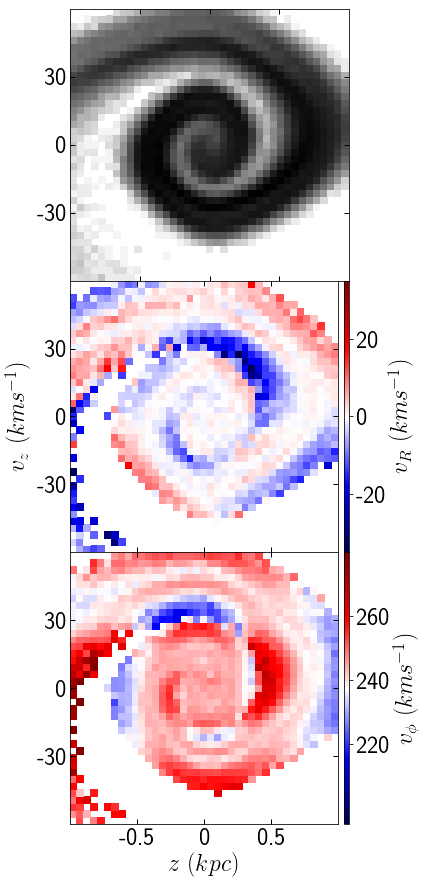}
	\caption{Same as Fig.\ \ref{ToyModelNoCoupling} but for the model
          considered in Section \ref{MWP} (realistic model for the
          Milky Way potential).}\label{MWTestParticle}
\end{figure}

The results are shown in Fig.\,\ref{MWTestParticle}.  The number
density spiral is a bit more tightly wound than our previous example
implying a stronger gradient in the vertical frequency with $E_z$.
More striking are the differences between the $v_R$ and $v_\phi$
spirals seen here and in our previous examples.  In particular, there
are now prominent gradients along the edges of the spirals.  For
example, along the inner edge of the $v_\phi$ spiral, the stars
alternate between high azimuthal velocity (red) and low azimuthal
velocity (blue) five times in a little over $360^\circ$.  The fact
that the variations in $v_\phi$ are strongest along the edges of the
spirals makes sense since these stars have the largest $E_z$ where the
effect of the coupling term is the largest.

\section{PHASE SPACE SPIRALS WITH SELF-GRAVITY}\label{selfgravity}

In the previous section, we considered test particles in a fixed
axisymmetric potential.  The evolution of their DF was therefore
governed by kinematic phase mixing.  Test particles were
used by \citet{delavega2015} in their study of bending and breathing
modes and by \citet{antoja2018} and \citet{binney2018} to explain the
{\it Gaia} spirals.  However, as stressed by \citet{hunter1969} (see also
\citet{sparke1988}) there are two effects associated with a bending
perturbation of a disc: the restoring force of the unperturbed disc on
the perturbation and the potential associated with the perturbation
acting on the unperturbed disc.  In effect, the test-particle models
treat the former and ignore the latter.  In linear perturbation
theory, both effects enter at the same order.  It follows that
self-gravity will be important for the evolution of bending waves and
phase space spirals.

In this Section we present results from simulations of a fully
self-consistent disc-bulge-halo system.  The initial conditions are
generated with \textsc{GalactICS}, which uses a disc DF given by
Eq.\,\ref{DFeq} with $g(L_z)$ chosen to yield an exponential profile
for the surface density (See \citet{kuijken1995, widrow2008}).  The
model parameters are chosen so that the rotation curve decomposition
into the three components matches the realistic Milky Way model
introduced in the previous section, as shown in Fig.\,\ref{vcirc}.  We
use 1M particles for the halo and 200K particles for the bulge.  To
boost mass resolution near the solar circle, we sample the region of
the disc between $6.5\,{\rm kpc}$ and $9.5\,{\rm kpc}$ with 5M
particles and the remainder with 1M particles.  The particles in the
ring then have a mass of about 1300 $M_\odot$ or about 28 times less
massive than the particles in the rest of the disc.

We introduce an {\it ad hoc} bend to the disc by applying a velocity
perturbation to both low and high resolution disc stars of the form
\begin{equation}
\delta v_z = v_0 e^{-\left (R - R_0\right )^2/2\delta_R^2}
\cos{\phi}
\end{equation}
\noindent where we choose $v_0=30\,{\rm km\,s^{-1}}$, $R_0 = 8\,{\rm
  kpc}$, and $\delta_R = 500\,{\rm pc}$.  Thus, initially one side of
the disc will bend to the North while the other will bend to the
South.

\begin{figure}
	\centering 
	\includegraphics[width=9cm]{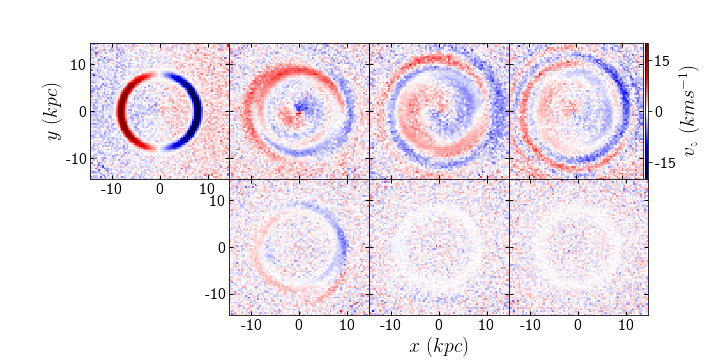}
	\caption{Face-on maps for the mean vertical velocity.  
          Left-most top panel shows the initial conditions for both  
self-consistent and test-particle simulations.  From left to right, 
one then has maps for the $250\,{\rm Myr}$, $500\,{\rm Myr}$, and  
$1\,{\rm Gyr}$ snapshots.  The top row shows results for the  
self-consistent simulation while the bottom row shows results for the  
test particle simulation.}\label{wmap}
\end{figure}

The system is evolved for $1\,{\rm Gyr}$ with a timestep of
  $10^5$ years and a Plummer softening of $50\,{\rm pc}$. Face-on maps
  of the mean vertical velocity for the $250\,{\rm Myr}$, $500\,{\rm
    Myr}$, and $1\,{\rm Gyr}$ snapshots are shown in
  Fig.\,\ref{wmap}. In making this figure we reoriented the disc so
  that the direction corresponding to the smallest eignevalue of the
  moment of inertia tensor and the angular momentum vector were normal
  to the plane of the map. Since the perturbation introduces a small
  angular momentum, when we reorient the disc, the region far from
  the perturbation is given a small $z$-velocity in the opposite sense
  of the perturbation, as seen, for example, in the upper-left panel.

To gain a handle on the effect self-gravity has on the evolution of
bending waves, we also follow the evolution of low and high
  resolution disc particles under the assumption that the background
potential is fixed and given by the unperturbed potential of the
model.  The face-on maps of the mean vertical velocity are shown in
Fig.\,\ref{wmap}.  We see in the bottom three panels that the
kinematic bending waves are rapidly damped, presumably due
to phase mixing.  By $500\,{\rm Myr}$, the vertical velocity
  signal is entirely due to particle noise and,
  as one can clearly see in the figure, is significantly less in the
  annulus dominated by low mass particles. Conversely, in the upper
  panels we see that the bending waves persist with roughly constant
  amplitude through to $ 1\ {\rm Gyr} $. Clearly, self-gravity is
  crucial to the persistence of bending waves in discs. 

We next search for phase space spirals in a manner similar to what was
done in \citet{antoja2018} with GDR2. In particular, we select stars
in a circular arc between $7.5\,{\rm kpc}$ and $8.5\,{\rm kpc}$ and
over a range in $\phi$ that spans $1\,{\rm rad}$.  This region
typically contains 250k particles, or about 1/20 of the total number
of particles in our high-resolution ring. The number of stars here is thus
a factor of $3-4$ less than the number of stars in the arcs considered
by \citet{antoja2018}. On the other hand, our arc is about five times
wider in $R$ and seven times wider in $\phi$ than the one they chose.
Thus, even with our ring of low mass particles, our simulations have poorer
particle and spatial resolution in comparison with the data.

\begin{figure}
	\centering 
	\includegraphics[width=9cm]{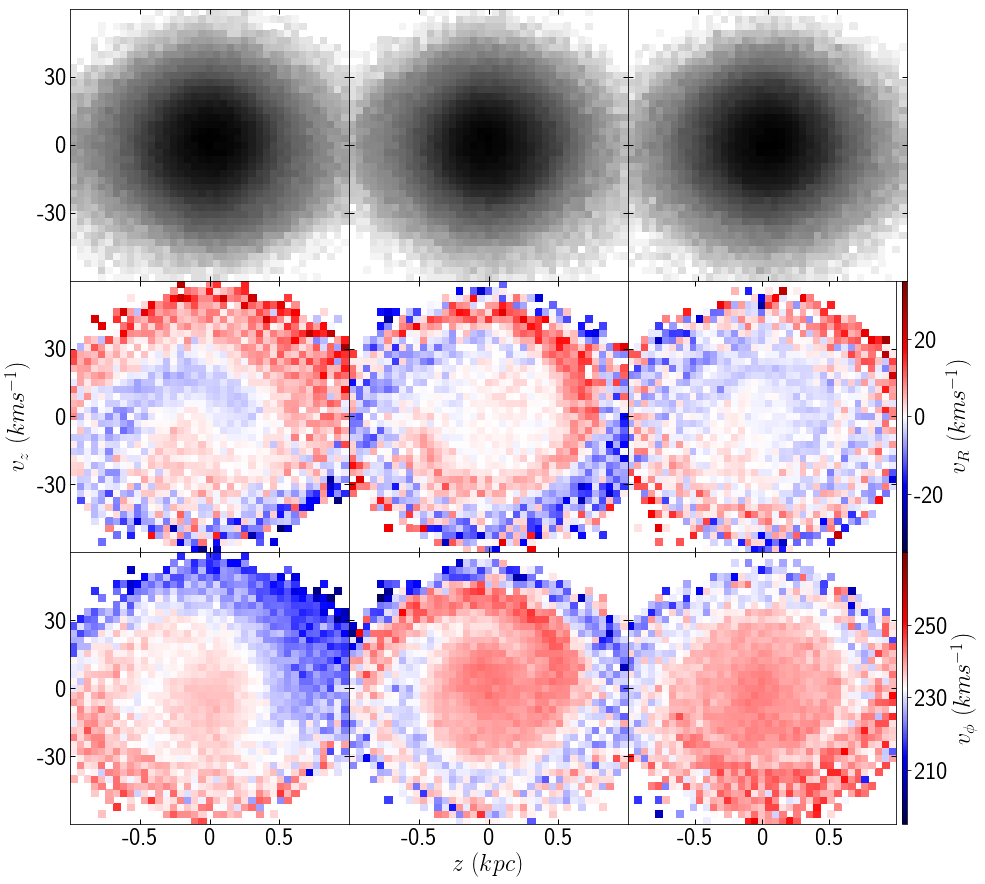}
	\caption{Number density and mean $v_R$ and $v_\phi$ maps for 
		the self-consistent simulation.  Each row has the same 
		format as in figures in Section\,\ref{KinematicSpirals}. 
		From left to right, the columns are for the 
		$250\,{\rm Myr}$, $500\,{\rm Myr}$, and 
		$1\,{\rm Gyr}$ snapshots.} 
	\label{SelfConsistent}
\end{figure}

\begin{figure}
	\centering 
	\includegraphics[width=9cm]{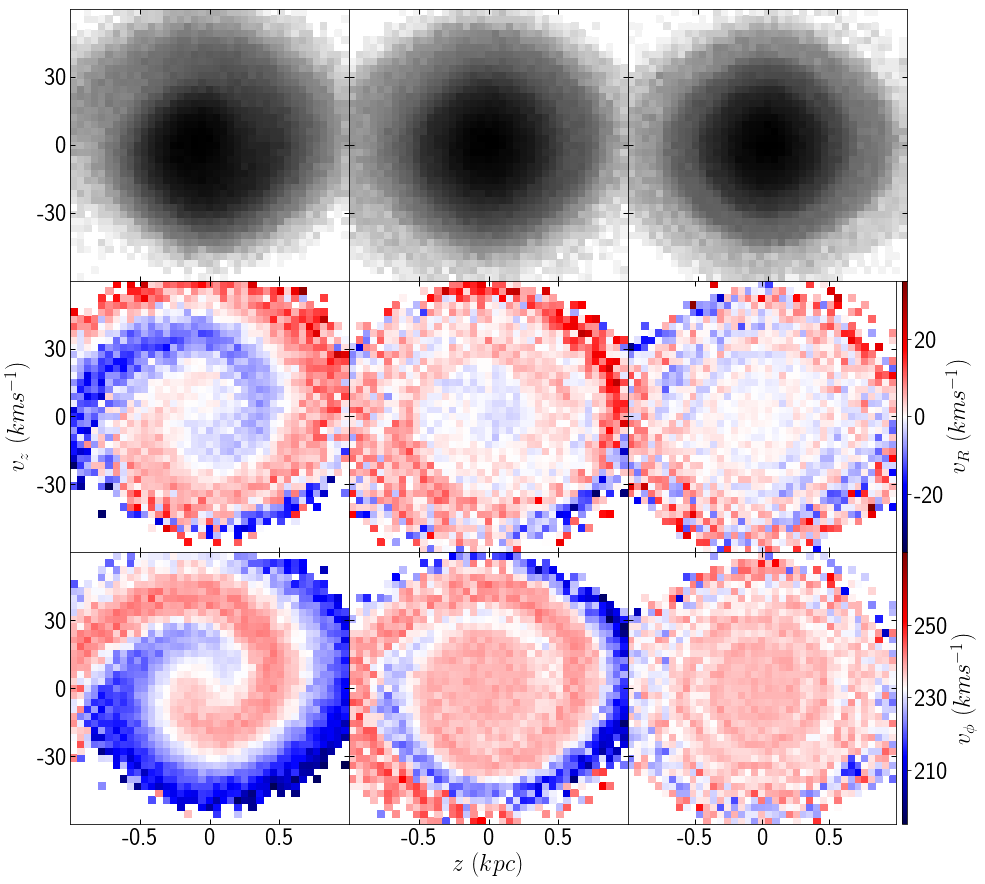}
	\caption{Same as Fig.\,\ref{SelfConsistent} but for the test
		particle simulation.}
	\label{TestParticle}
\end{figure}

 The number counts and mean $v_r$ and $v_\phi$ across the
  $z-v_z$ plane are shown in Fig.~\ref{SelfConsistent} and
  Fig.~\ref{TestParticle} for the same three snapshots that were used in
  Fig.\,\ref{wmap}. Spirals are found in both mean $v_R$ and $v_\phi$
  with an amplitude of $\sim 10-20\,{\rm km\,s^{-1}}$, which is
  comparable to, though somewhat smaller than, the amplitude of the
  initial perturbation.  As expected, the pattern winds up over time.
  In the $v_\phi$ panel of Fig.\,1 from \citet{antoja2018}, one
  observes roughly one and a half wraps within $|z|< 700\,{\rm
    pc}$. Our self consistent simulation appears to reach a comparable
  pattern around $t\simeq 1\,{\rm Gyr}$, though by this time, the
  pattern is becoming difficult to discern.

The phase space spirals are more distinct in the test particle run.
Indeed, a spiral pattern can been easily seen in the number counts,
which is not the case in the self-consistent simulation. The
degree of winding seen in the self-consistent and test-particle runs
is fairly similar through the $500\,{\rm Myr}$ snapshot. However,
by $1\,{\rm Gyr}$, the test-particle run shows a tightly wound
spiral with four phase space wrappings within $|z|<1\,{\rm kpc}$
while the self-consistent run shows about half that many.

The patterns seen in Figs.\,\ref{TestParticle} and
\ref{SelfConsistent} as well as Fig.\,\ref{wmap} can be understood as
follows. In our test particle simulation, the evolution of the phase
space DF is driven entirely by phase mixing in an analytic,
time-independent, axisymmetric potential. It is therefore not
surprising that over time, the DF develops intricate, fine-grained
patterns such as those seen in the $1\,{\rm Gyr}$ panel of
Fig.\,\ref{TestParticle}. These patterns are typically washed out when
one maps moments of the DF, as in Fig.\,\ref{wmap}. By contrast, when
self-gravity is included, the disc develops persistent vertical
oscillations, as seen in Fig\,\ref{wmap}, that is roughly constant
between $500\,{\rm Myr}$ and $1\,{\rm Gyr}$. The spiral pattern seen
in the $1\,{\rm Gyr}$ panel of Fig.\,\ref{SelfConsistent} may well be
a manifestation of these persistent oscillations.

Two effects may explain why the spirals in our self-consistent
simulation are less well-defined than in the test-particle
case. First, the disc in the self-consistent simulation can develop
non-axisymmetric features, such as spiral structures, in addition to
the features that arise from the initial perturbation. Thus, the
particles that end up in the circular arc used in
Fig.\,\ref{SelfConsistent} will have been subjected to a more
complicated forcing function than those in the test-particle
run. Second, particles in the self-consistent run experience two-body
relaxation effects not present in our test-particle
simulation. However, the two-body relaxation time for the disc is
longer than the duration of the simulation by several orders of
magnitude. We also note that additional simulations indicate that our
results are relatively insensitive to choice of softening length and
time-step. Nevertheless, no N-body simulation is perfectly
collisionless. The combination of these effects may explain why the
spiral patterns that arise when including self gravity are not as well-defined as those with test particles.

\section{CONCLUSIONS}

In this paper we have focused on three facets of phase space spirals.
First, spirals similar to those found in {\it Gaia} DR2 naturally arise when the disc experiences a bending 
perturbation.  In general, a passing satellite will cause the disc
to bend \citep{toth1992, sellwood1998, widrow2014, binney2018},
and within a few dynamical times these bends lead to spirals 
in the $z-v_z-v_R-v_\phi$ space.  Second, the non-separable nature
of the effective potential is crucial for understanding the morphology
of the spirals.  This effect, which appears at cubic and higher 
order in the Taylor expansion of the effective potential, implies a
coupling of the in-plane and vertical epicyclic motions and leads to
variations in $v_R$ and $v_\phi$ across and along the edges of the
spirals.  Finally, self gravity appears to be an essential ingredient 
in any study of bending waves in discs.

The results from our toy-model simulations suggest that by studying
the morphology of phase space spirals, we can probe the detailed
structure of the Galactic potential.  In particular, one might gain a
handle on the coefficients of cubic and quartic terms in the Taylor
expansion of the effective potential near the Sun and beyond, once
future data releases from {\it Gaia} become available.

The natural way forward is to test these ideas with different models
for the Galactic potential.  Studies of this type will allow us to
determine just how sensitive the phase space spirals are to the
potential.  Such a program should be straightforward to accomplish
with kinematic models where test particles are used as probes.
However, our self-consistent simulations suggest that these models may
miss the essential physics of self-gravity in the perturbed disc.
Unfortunately, self-consistent simulations are computationally expensive, and the mass and spatial resolution that can be achieved, especially if one aims to study a large
parameter space, are well below the resolution available in the data.

\section*{Acknowledgments} {We thank Jo Bovy for help with the
  Galactic potential used in Section 2.  This works was supported by a
  Discovery Grant with the Natural Sciences and Engineering Research
  Council of Canada.}


\bibliographystyle{mnras}
\bibliography{bibliography_Nov13.bib} 

\begin{thebibliography}{}
\makeatletter
\relax
\def\mn@urlcharsother{\let\do\@makeother \do\$\do\&\do\#\do\^\do\_\do\%\do\~}
\def\mn@doi{\begingroup\mn@urlcharsother \@ifnextchar [ {\mn@doi@}
  {\mn@doi@[]}}
\def\mn@doi@[#1]#2{\def\@tempa{#1}\ifx\@tempa\@empty \href
  {http://dx.doi.org/#2} {doi:#2}\else \href {http://dx.doi.org/#2} {#1}\fi
  \endgroup}
\def\mn@eprint#1#2{\mn@eprint@#1:#2::\@nil}
\def\mn@eprint@arXiv#1{\href {http://arxiv.org/abs/#1} {{\tt arXiv:#1}}}
\def\mn@eprint@dblp#1{\href {http://dblp.uni-trier.de/rec/bibtex/#1.xml}
  {dblp:#1}}
\def\mn@eprint@#1:#2:#3:#4\@nil{\def\@tempa {#1}\def\@tempb {#2}\def\@tempc
  {#3}\ifx \@tempc \@empty \let \@tempc \@tempb \let \@tempb \@tempa \fi \ifx
  \@tempb \@empty \def\@tempb {arXiv}\fi \@ifundefined
  {mn@eprint@\@tempb}{\@tempb:\@tempc}{\expandafter \expandafter \csname
  mn@eprint@\@tempb\endcsname \expandafter{\@tempc}}}

\bibitem[\protect\citeauthoryear{{Antoja} et~al.,}{{Antoja}
  et~al.}{2018}]{antoja2018}
{Antoja} T.,  et~al., 2018, preprint (\mn@eprint {arXiv} {1804.10196})

\bibitem[\protect\citeauthoryear{{Bennett} \& {Bovy}}{{Bennett} \&
  {Bovy}}{2018}]{bennett2018}
{Bennett} M.,  {Bovy} J.,  2018, \mn@doi [\mnras] {10.1093/mnras/sty2813}, 482,
  1417

\bibitem[\protect\citeauthoryear{{Binney}}{{Binney}}{1992}]{binney1992}
{Binney} J.,  1992, \mn@doi [\araa] {10.1146/annurev.aa.30.090192.000411}, 30,
  51

\bibitem[\protect\citeauthoryear{{Binney} \& {Schoenrich}}{{Binney} \&
  {Schoenrich}}{2018}]{binney2018}
{Binney} J.,  {Schoenrich} R.,  2018, preprint (\mn@eprint {arXiv}
  {1807.09819})

\bibitem[\protect\citeauthoryear{{Binney} \& {Tremaine}}{{Binney} \&
  {Tremaine}}{2008}]{binney2008}
{Binney} J.,  {Tremaine} S.,  2008, {Galactic Dynamics: Second Edition}.
Princeton University Press

\bibitem[\protect\citeauthoryear{{Bovy}}{{Bovy}}{2015}]{bovy2015}
{Bovy} J.,  2015, \mn@doi [\apjs] {10.1088/0067-0049/216/2/29}, 216, 29

\bibitem[\protect\citeauthoryear{{Carlin} et~al.,}{{Carlin}
  et~al.}{2013}]{carlin2013}
{Carlin} J.~L.,  et~al., 2013, \mn@doi [\apjl] {10.1088/2041-8205/777/1/L5},
  777, L5

\bibitem[\protect\citeauthoryear{{Chequers} \& {Widrow}}{{Chequers} \&
  {Widrow}}{2017}]{chequers2017}
{Chequers} M.~H.,  {Widrow} L.~M.,  2017, \mn@doi [\mnras]
  {10.1093/mnras/stx2165}, 472, 2751

\bibitem[\protect\citeauthoryear{{Chequers}, {Widrow}  \& {Darling}}{{Chequers}
  et~al.}{2018}]{chequers2018}
{Chequers} M.~H.,  {Widrow} L.~M.,   {Darling} K.,  2018, preprint (\mn@eprint
  {arXiv} {1805.12449})

\bibitem[\protect\citeauthoryear{{Debattista}}{{Debattista}}{2014}]{debattista2014}
{Debattista} V.~P.,  2014, \mn@doi [\mnras] {10.1093/mnrasl/slu069}, 443, L1

\bibitem[\protect\citeauthoryear{{Feldmann} \& {Spolyar}}{{Feldmann} \&
  {Spolyar}}{2015}]{feldman2015}
{Feldmann} R.,  {Spolyar} D.,  2015, \mn@doi [\mnras] {10.1093/mnras/stu2147},
  446, 1000

\bibitem[\protect\citeauthoryear{{Gaia Collaboration} et~al.,}{{Gaia
  Collaboration} et~al.}{2018b}]{katz2018}
{Gaia Collaboration} et~al., 2018b, preprint (\mn@eprint {arXiv} {1804.09380})

\bibitem[\protect\citeauthoryear{{Gaia Collaboration}, {Brown}, {Vallenari},
  {Prusti}, {de Bruijne}, {Babusiaux}  \& {Bailer-Jones}}{{Gaia Collaboration}
  et~al.}{2018a}]{brown2018}
{Gaia Collaboration} {Brown} A.~G.~A.,  {Vallenari} A.,  {Prusti} T.,  {de
  Bruijne} J.~H.~J.,  {Babusiaux} C.,   {Bailer-Jones} C.~A.~L.,  2018a,
  preprint (\mn@eprint {arXiv} {1804.09365})

\bibitem[\protect\citeauthoryear{{G{\'o}mez}, {Minchev}, {O'Shea}, {Beers},
  {Bullock}  \& {Purcell}}{{G{\'o}mez} et~al.}{2013}]{gomez2013}
{G{\'o}mez} F.~A.,  {Minchev} I.,  {O'Shea} B.~W.,  {Beers} T.~C.,  {Bullock}
  J.~S.,   {Purcell} C.~W.,  2013, \mn@doi [\mnras] {10.1093/mnras/sts327},
  429, 159

\bibitem[\protect\citeauthoryear{{G{\'o}mez}, {White}, {Grand}, {Marinacci},
  {Springel}  \& {Pakmor}}{{G{\'o}mez} et~al.}{2017}]{gomez2017}
{G{\'o}mez} F.~A.,  {White} S.~D.~M.,  {Grand} R.~J.~J.,  {Marinacci} F.,
  {Springel} V.,   {Pakmor} R.,  2017, \mn@doi [\mnras]
  {10.1093/mnras/stw2957}, 465, 3446

\bibitem[\protect\citeauthoryear{{Hernquist}}{{Hernquist}}{1990}]{hernquist1990}
{Hernquist} L.,  1990, \mn@doi [\apj] {10.1086/168845}, 356, 359

\bibitem[\protect\citeauthoryear{{Hunter} \& {Toomre}}{{Hunter} \&
  {Toomre}}{1969}]{hunter1969}
{Hunter} C.,  {Toomre} A.,  1969, {Dynamics of the Bending of the Galaxy}.
p.~747, \mn@doi{10.1086/149908}

\bibitem[\protect\citeauthoryear{{Kuijken} \& {Dubinski}}{{Kuijken} \&
  {Dubinski}}{1995}]{kuijken1995}
{Kuijken} K.,  {Dubinski} J.,  1995, \mn@doi [\mnras]
  {10.1093/mnras/277.4.1341}, 277, 1341

\bibitem[\protect\citeauthoryear{{Kuijken} \& {Gilmore}}{{Kuijken} \&
  {Gilmore}}{1989}]{kuijken1989}
{Kuijken} K.,  {Gilmore} G.,  1989, \mn@doi [\mnras] {10.1093/mnras/239.2.571},
  239, 571

\bibitem[\protect\citeauthoryear{{Miyamoto} \& {Nagai}}{{Miyamoto} \&
  {Nagai}}{1975}]{miyamoto1975}
{Miyamoto} M.,  {Nagai} R.,  1975, \pasj, 27, 533

\bibitem[\protect\citeauthoryear{{Monari}, {Famaey}  \& {Siebert}}{{Monari}
  et~al.}{2015}]{monari2015}
{Monari} G.,  {Famaey} B.,   {Siebert} A.,  2015, \mn@doi [\mnras]
  {10.1093/mnras/stv1206}, 452, 747

\bibitem[\protect\citeauthoryear{{Navarro}, {Frenk}  \& {White}}{{Navarro}
  et~al.}{1997}]{navarro1997}
{Navarro} J.~F.,  {Frenk} C.~S.,   {White} S.~D.~M.,  1997, \mn@doi [\apj]
  {10.1086/304888}, 490, 493

\bibitem[\protect\citeauthoryear{{Sch{\"o}nrich} \& {Dehnen}}{{Sch{\"o}nrich}
  \& {Dehnen}}{2018}]{schonrich2015}
{Sch{\"o}nrich} R.,  {Dehnen} W.,  2018, \mn@doi [\mnras]
  {10.1093/mnras/sty1256}, 478, 3809

\bibitem[\protect\citeauthoryear{{Sellwood}}{{Sellwood}}{2013}]{sellwood2013}
{Sellwood} J.~A.,  2013, {Dynamics of Disks and Warps}.
p.~923, \mn@doi{10.1007/978-94-007-5612-0_18}

\bibitem[\protect\citeauthoryear{{Sellwood}, {Nelson}  \&
  {Tremaine}}{{Sellwood} et~al.}{1998}]{sellwood1998}
{Sellwood} J.~A.,  {Nelson} R.~W.,   {Tremaine} S.,  1998, \mn@doi [\apj]
  {10.1086/306280}, 506, 590

\bibitem[\protect\citeauthoryear{{Smith}, {Flynn}, {Candlish}, {Fellhauer}  \&
  {Gibson}}{{Smith} et~al.}{2015}]{smith2015}
{Smith} R.,  {Flynn} C.,  {Candlish} G.~N.,  {Fellhauer} M.,   {Gibson} B.~K.,
  2015, \mn@doi [\mnras] {10.1093/mnras/stv228}, 448, 2934

\bibitem[\protect\citeauthoryear{{Sparke} \& {Casertano}}{{Sparke} \&
  {Casertano}}{1988}]{sparke1988}
{Sparke} L.~S.,  {Casertano} S.,  1988, \mn@doi [\mnras]
  {10.1093/mnras/234.4.873}, 234, 873

\bibitem[\protect\citeauthoryear{{Toth} \& {Ostriker}}{{Toth} \&
  {Ostriker}}{1992}]{toth1992}
{Toth} G.,  {Ostriker} J.~P.,  1992, \mn@doi [\apj] {10.1086/171185}, 389, 5

\bibitem[\protect\citeauthoryear{{Widrow}, {Pym}  \& {Dubinski}}{{Widrow}
  et~al.}{2008}]{widrow2008}
{Widrow} L.~M.,  {Pym} B.,   {Dubinski} J.,  2008, \mn@doi [\apj]
  {10.1086/587636}, 679, 1239

\bibitem[\protect\citeauthoryear{{Widrow}, {Gardner}, {Yanny}, {Dodelson}  \&
  {Chen}}{{Widrow} et~al.}{2012}]{widrow2012}
{Widrow} L.~M.,  {Gardner} S.,  {Yanny} B.,  {Dodelson} S.,   {Chen} H.-Y.,
  2012, \mn@doi [\apjl] {10.1088/2041-8205/750/2/L41}, 750, L41

\bibitem[\protect\citeauthoryear{{Widrow}, {Barber}, {Chequers}  \&
  {Cheng}}{{Widrow} et~al.}{2014}]{widrow2014}
{Widrow} L.~M.,  {Barber} J.,  {Chequers} M.~H.,   {Cheng} E.,  2014, \mn@doi
  [\mnras] {10.1093/mnras/stu396}, 440, 1971

\bibitem[\protect\citeauthoryear{{Williams} et~al.,}{{Williams}
  et~al.}{2013}]{williams2013}
{Williams} M.~E.~K.,  et~al., 2013, \mn@doi [\mnras] {10.1093/mnras/stt1522},
  436, 101

\bibitem[\protect\citeauthoryear{{Xu}, {Newberg}, {Carlin}, {Liu}, {Deng},
  {Li}, {Sch{\"o}nrich}  \& {Yanny}}{{Xu} et~al.}{2015}]{xu2015}
{Xu} Y.,  {Newberg} H.~J.,  {Carlin} J.~L.,  {Liu} C.,  {Deng} L.,  {Li} J.,
  {Sch{\"o}nrich} R.,   {Yanny} B.,  2015, \mn@doi [\apj]
  {10.1088/0004-637X/801/2/105}, 801, 105

\bibitem[\protect\citeauthoryear{{Yanny} \& {Gardner}}{{Yanny} \&
  {Gardner}}{2013}]{yanny2013}
{Yanny} B.,  {Gardner} S.,  2013, \mn@doi [\apj] {10.1088/0004-637X/777/2/91},
  777, 91

\bibitem[\protect\citeauthoryear{{de la Vega}, {Quillen}, {Carlin},
  {Chakrabarti}  \& {D'Onghia}}{{de la Vega} et~al.}{2015}]{delavega2015}
{de la Vega} A.,  {Quillen} A.~C.,  {Carlin} J.~L.,  {Chakrabarti} S.,
  {D'Onghia} E.,  2015, \mn@doi [\mnras] {10.1093/mnras/stv2055}, 454, 933

\makeatother
\end{thebibliography}


\bsp    
\label{lastpage}

\end{document}